\documentclass[11pt]{article}
\usepackage{times} 
\usepackage{natbib}
\usepackage{geometry}
\usepackage[small,bf]{caption}
\usepackage{graphicx}
\usepackage{amsbsy}
\usepackage{amsmath,amssymb}
\usepackage{amsthm}
\usepackage{url}
\usepackage{amsfonts}
\usepackage{lscape}
\usepackage{verbatim}
\usepackage{subfigure}
\usepackage{dsfont}
\usepackage{chapterbib}
\usepackage{color}
\usepackage{multirow}
\usepackage{bm}
\usepackage{authblk}

\usepackage[breaklinks]{hyperref}
\hypersetup{pdfborder={0 0 0},
	colorlinks=true,
	linkcolor=black,
	citecolor=blue,
	urlcolor=black,
	hyperindex=true}

\DeclareMathSymbol{\Real}{\mathbin}{AMSb}{"52}
\def\pr{{\rm Pr}}

\def\E{{\rm E}}
\def\calP{{\mathcal{P}}}

\title{Efficient inference and simulation for elliptical Pareto processes}

\author[1]{Emeric Thibaud\footnote{Emeric.Thibaud@colostate.edu}}
\author[2]{Thomas Opitz\footnote{Thomas.Opitz@paca.inra.fr}}

\affil[1]{Department of Statistics, Colorado State University, Fort Collins, Colorado 80523-1877, U.S.A.}
\affil[2]{Biostatistics and Spatial Processes Unit, French National Institute for Agronomic Research, 84914 Avignon, France}

\date{}

\begin{document}

\maketitle

\begin{abstract}
Recent advances in extreme value theory have established $\ell$-Pareto processes as the natural limits for extreme events defined in terms of exceedances of a risk functional. Here we provide methods for the practical modelling of data based on a tractable yet flexible dependence model. We introduce the class of elliptical $\ell$-Pareto processes, which arise as the limit of threshold exceedances of certain elliptical processes characterized by a correlation function and a shape parameter. An efficient inference method based on maximizing a full likelihood with partial censoring is developed. Novel procedures for exact conditional and unconditional simulation are proposed. These ideas are illustrated using precipitation extremes in Switzerland.
\bigskip


\end{abstract}

%%%%%%%%%%%%%%%%%%%%%%%%%%%%%%%%%%%%%%%%%%%%%%%%%%%%%%%%%%%%%%%%

\section{Introduction}
There has recently been increasing development of methodologies for modelling spatial extremes, motivated by numerous applications in climatology and environmental sciences. Classical extreme value theory relies on max-stable processes, which extend the univariate generalized extreme-value distribution to stochastic processes. Such processes are the only possible nondegenerate limits for rescaled maxima of spatial processes \citep[][Ch.~9]{deHaan.Ferreira:2006} and provide a natural modelling framework for asymptotically dependent extremes. 

Inference for spatial extremes has been based on various max-stable models \citep[e.g.,][]{Davison.etal:2012,Ribatet:2013}. The extremal Gaussian model \citep{Schlather:2002} or the Brown--Resnick model \citep{Kabluchko.etal:2009} have proven to be well-suited for modelling extremal dependence of environmental data \citep{Davison.etal:2012,Ribatet:2013}. The extremal-$t$ process, which can be seen as generalizing these two models, is the max-stable limit of all asymptotically dependent elliptical processes \citep{Opitz:2013}. Its distribution depends on a correlation function and a shape parameter, providing a flexible dependence structure for spatial extremes. Because of the complicated form of the distribution of a max-stable process, composite likelihood methods have been used to fit such models \citep{Padoan.etal:2010}, leading to a loss in efficiency. More recently, efficient full likelihood inference methods were developed in a point process framework. \citet{Engelke.etal:2012} developed full likelihood methods based either on the distribution of increments with respect to a fixed extreme component, or on the multivariate spectral measure. \citet{Wadsworth.Tawn:2013} calculated a full likelihood for exceedances of a thresholding field while censoring the part of the observation vector falling below this threshold. 

In this paper, we propose the use of $\ell$-Pareto processes \citep{Dombry.Ribatet:2013} for modelling extremes of spatial processes defined in terms of the exceedance  of a risk functional. \citet{Ferreira.deHaan:2012} and \citet{Dombry.Ribatet:2013} showed that Pareto processes are the only possible asymptotic limits for threshold exceedances of spatial processes. Inference based on these processes is currently limited to nonparametric estimation \citep{Dombry.Ribatet:2013}. We introduce the elliptical $\ell$-Pareto process, which is the limiting process for threshold exceedances of all asymptotically dependent elliptical processes, and propose an efficient inference approach for it based on a full likelihood with partial censoring. The resulting inferential procedures, potentially more efficient than composite likelihood methods, are discussed, and efficiency gains over a pairwise likelihood are assessed in a simulation study. In addition, we propose a new approach to exact simulation from extremal-$t$ and elliptical Pareto processes, and we show how conditional simulations can be obtained very easily for the latter. Finally, we illustrate the use of elliptical Pareto processes in an application to extreme precipitation in Switzerland.

We develop our results for processes with continuous sample paths defined on a nonempty compact domain $K\subset \mathbb{R}^m$, $m\geq 1$. The assumption of continuity is natural in applications and ensures that Pareto processes are well-defined. We focus here on the practical use of Pareto processes; for more technical details on the definitions of these processes and related convergence in functional spaces we refer to \citet{Ferreira.deHaan:2012} and \citet{Dombry.Ribatet:2013}.

\section{Functional extreme value theory}\label{sec:funcevt}

\subsection{$\ell$-Pareto processes}

We let $C(K)$ denote the space of continuous functions over $K$, endowed with the supremum norm $\| f \|_\infty=\sup_{s\in K} |f(s)|$. The restriction of $C(K)$ to non-negative functions is denoted by $C_+(K)$. In univariate and multivariate theory, a generalized Pareto limit is obtained by conditioning on the exceedance of a high threshold in at least one component \citep{Rootzen.Tajvidi:2006}. \citet{Ferreira.deHaan:2012} extended this idea to infinite-dimensional spaces by conditioning on exceedances of the supremum of the process over the space, leading to the notion of a generalized Pareto process. This definition was generalized by \citet{Dombry.Ribatet:2013} to so-called $\ell$-Pareto processes by considering exceedances defined in terms of a linear risk functional.

A functional $\ell:C_+(K)\rightarrow [0,\infty)$ is called a risk functional, or cost functional, if it is continuous and homogeneous, i.e., $\ell(t f) = t\,\ell( f)$ for $t\geq 0$. In what follows, the unit sphere in $C_+(K)$ with respect to $\ell$ is written $S_\ell=\{ f \in C_+(K) : \ell( f)=1\}$. Suppose that we are given  a risk functional $\ell$ and  a probability measure $\rho$ on $S_\ell$. We call any process $ Y^* = \{Y^*(s)\}_{s\in K}$  a standard $\ell$-Pareto process with $\ell$-spectral distribution $\rho$ if it can be represented as
\begin{equation}\label{eq:constrprocpareto}
Y^*(s)= R f_0(s),\quad R\sim \mathrm{Par}(1), \quad \{f_0(s)\}\sim \rho,
\end{equation}
i.e., $\pr(R>y)=1/y$ ($y\geq1$), with $R$ independent of the spectral function $ f_0=\{f_0(s)\}$.
For continuous real functions $\sigma(s)> 0$, $\mu(s)$, $\xi(s)$ defined over  $K$, the process 
\begin{equation}\label{eq:trafoppclassgen}
\left\{
\begin{array}{ll}
\mu(s) + \sigma(s) \left\{Y^*(s)^{\xi(s)}-1\right\}/\xi(s) , \quad &  \xi(s)\neq 0,\\
\mu(s) + \sigma(s) \log Y^*(s),\quad& \xi(s)=0,
\\
\end{array}
\right. \quad s \in K,
\end{equation}
is termed a generalized $\ell$-Pareto process. To avoid confusion between processes associated to different risk functionals $\ell$, we will write $Y^*_\ell$ for $Y^*$. Such processes $Y^*_\ell$ generalize the peaks-over-threshold stability of multivariate generalized Pareto distributions to infinite dimensions: for any $u\geq 1$, the renormalized threshold-exceeding process $\left\{u^{-1}Y^*_\ell \mid \ell(Y^*_\ell) \geq u\right\}$ is equal to $Y^*_\ell$ in distribution \citep{Dombry.Ribatet:2013}. The interpretation of the construction~\eqref{eq:constrprocpareto} is that $R$ characterizes the intensity of an extreme event in terms of the risk functional $\ell$, whereas $f_0$ describes the corresponding spatial profile. 

In applications, different choices of $\ell$ may be used to answer different questions. If $\ell( f)=\max_{j=1,\ldots,D} f(s_j)/u_j$ for certain sites $s_j\in K$ $(j=1,\ldots,D)$, we focus on processes with at least one exceedance of the thresholds $u_j>0$. By contrast, $\ell( f)=\min_{j=1,\ldots,D} f(s_j)/u_j$ requires exceedances at each of the $D$ sites. The original definition of a Pareto process \citep{Ferreira.deHaan:2012} uses $\ell(f)=\sup_{s\in K} f(s)$, but conditioning on another $\ell(f)$ is desirable in applications where data are only observed at a finite number of sites.

\subsection{Limiting processes of extremes}
We recall the different forms of convergence of extremes of continuous processes in terms of block maxima, threshold exceedances and point processes. Throughout, the symbol $\Longrightarrow$ indicates weak convergence of random elements from the univariate, multivariate or functional domain. 
 For independent and identically distributed copies $X_1,X_2,\ldots$ of a  stochastic process $X = \{X(s)\}_{s\in K}$  with continuous sample paths, we say that $X$ is in the maximum domain of attraction of a max-stable process $Z = \{Z(s)\}_{s\in K}$ \citep[][Ch.~9]{deHaan.Ferreira:2006} if there exist sequences of normalizing continuous functions $a_n(s)>0$ and $b_n(s)$ such that 
\begin{equation}\label{eq:convms}
\left\{ \max_{i=1,\ldots,n} a_n(s)^{-1}\{X_i(s)-b_n(s)\}  \right\}  \Longrightarrow \left\{Z(s)\right\}, \quad n\rightarrow \infty, 
\end{equation}
in $C(K)$, with the limit process $Z$ having nondegenerate univariate distributions. Convergence of the dependence structure and of marginal distributions can be viewed separately in~\eqref{eq:convms} \citep[][\S9.2]{deHaan.Ferreira:2006}. Therefore, we define a normalized process $X^*$ by $X^*(s)=1/[1-F_{X(s)}\{X(s)\}]$ where $F_{X(s)}$ denotes the distribution of $X(s)$.  
If we assume that $X$ has continuous marginal distributions, then $X^*$ has marginal standard Pareto distributions. 
For $a_n(s)\equiv n$ and $b_n(s)\equiv 0$ the max-stable limit for $X^*$ in~\eqref{eq:convms} is a standard max-stable process $Z^*$ with univariate unit Fr\'echet distributions.

The distribution of the process $Z^*$ is fully characterized by a so-called exponent measure $\Lambda$ on $C(K)\setminus\{0\}$ through the relation \citep{Gine.etal:1990}
\begin{equation}\label{eq:charlambda}
\Lambda\left[\bigcup_{j=1,\ldots,D} \left\{ f\in C(K) : \sup_{s\in K_j} f(s) \geq z_j\right\} \right]   =   -\log \pr\left\{\sup_{s\in K_1} Z^*(s) \leq z_1, \ldots, \sup_{s\in K_D} Z^*(s) \leq z_D\right\}
\end{equation}
for any collection of nonempty compact sets $K_j\subset K$ and $z_j > 0$ $(j=1,\ldots,D)$. The measure $\Lambda$  is uniquely defined if we impose the constraint $\Lambda\{C (K)\setminus C_+(K)\}=0$. We obtain the unique version as $\Lambda_+(B)=\Lambda[\{  f \in C(K) :  f_+\in B\}]$, for measurable $B\subset C_+(K)\setminus\{0\}$, with $f_+(s)=\max\{f(s),0\}$. When the sets $K_j=\{s_j\}$ are singletons, expression~\eqref{eq:charlambda} is called the exponent function and denoted by $V(z_1,\ldots,z_D)$, where $V(1,\ldots,1)$ is known as the extremal coefficient of the sites $s_1,\ldots,s_D$. Finite-dimensional marginal measures of $\Lambda$ relative to $D$ sites $s=(s_1,\ldots,s_D)$ are written as $\Lambda_{s}$, i.e., $\Lambda_{s}(\times_{j=1,\ldots,D} [a_j,b_j]) = \Lambda[\bigcap_{j=1,\ldots,D}\{ f\in C(K) :  f(s_j)\in [a_j,b_j]\}]$ for $0<a_j<b_j$ $(j=1,...,D)$. In particular,  $V(z_1,\ldots,z_D)=\Lambda_{s}\left\{(\times_{j=1,\ldots,D} [-\infty,z_j])^C\right\}$.  

Max-stability of $Z^*$ implies that the measure $\Lambda_+$ is homogeneous of order $-1$, i.e., $\Lambda_+(tB)=t^{-1}\Lambda_+(B)$, $t>0$. For some risk functional $\ell$ and $f\in C_+(K)$ with $\ell(f)>0$, consider the pseudo-polar coordinates $(r, f_0)$ with $r=\ell(f)$ and $f_0=f/\ell( f)$. If $\kappa_\ell(K)=\Lambda_+\{\ell( f) \geq 1\}>0$, arguments similar to \citet[][\S9.4]{deHaan.Ferreira:2006} imply the factorization
\begin{equation}
\label{eq:lambdapolar}
\Lambda_+(\mathrm{d}f) = \kappa_\ell(K)\,  r^{-2}\mathrm{d}r \rho_\ell(\mathrm{d}  f_0), \quad  r>0, 
\end{equation}
with $\rho_\ell$ an $\ell$-spectral distribution on $S_\ell$. 

Assuming $X^*$ is in the maximum domain of attraction of a standard max-stable process $Z^*$, \citet[Theorem 3]{Dombry.Ribatet:2013} proved the convergence of standard $\ell$-exceedances
\begin{equation}\label{eq:convellPareto}
   \left\{  n^{-1} X^*(s) : \ell(X^*) > n \right\} \Longrightarrow  \{ Y^*_\ell(s)\},\quad n\rightarrow \infty,
\end{equation}
where $Y^*_\ell$ is a standard $\ell$-Pareto process with $\ell$-spectral distribution $\rho_\ell$ related to the exponent measure $\Lambda_+$ of $Z^*$ through~\eqref{eq:lambdapolar}. 

Convergence~\eqref{eq:convellPareto} establishes a basis for threshold-based inference as follows. From definition~\eqref{eq:constrprocpareto}, the distribution of a standard $\ell$-Pareto process is $r^{-2}\mathrm{d}r \rho_\ell(\mathrm{d}  f_0)$ on $[1,\infty)\times S_\ell$, which is also equal to $\Lambda_+(\mathrm{d}f)/\kappa_\ell(K)$ from~\eqref{eq:lambdapolar} with $f=rf_0$.  Hence the convergence in~\eqref{eq:convellPareto} conveys that, for large $n$,
\begin{equation*}
\pr\left\{X^* \in B \mid \ell(X^*) > n \right\} \approx n\Lambda_+(B)/\kappa_\ell(K),
\end{equation*}
for $B\subset\{f\in C_+(K): \ell(f)>n\}$.

A point process framework links max-stable limits for maxima and $\ell$-Pareto limits for threshold exceedances. Convergence~\eqref{eq:convms} for the normalized process $X^*$  is equivalently expressed in terms of the standard point process convergence \citep[][Theorem~9.3.9]{deHaan.Ferreira:2006}
\begin{equation}\label{eq:convPP}
   \left\{  n^{-1} X_i^*(s), i=1,\ldots,n\right\}  \Longrightarrow \calP, \quad n\rightarrow \infty, 
\end{equation}
where $\calP=\{ P_i(s),  i=1,2, \ldots\}$ is a Poisson process with  intensity measure $\Lambda_+$. Then, $Z^*(s)=\max_{i=1,\ldots,n} P_i(s)$ and, from Poisson process theory \citep[][Ch.~9]{Daley.Vere-Jones:2007}, the points $P_i$ with $\ell(P_i)\geq 1$ are independent and have distribution $\Lambda_+(\mathrm{d}f)/\kappa_\ell(K)$; they are realisations of the $\ell$-Pareto process $Y^*_\ell$.

Max-stable convergence in~\eqref{eq:convms} implies the convergence of marginal pointwise maxima to generalized extreme-value distributions, for which standard theory \citep[][Ch.~5]{Beirlant.etal:2004} provides alternative characterizations in terms of univariate threshold exceedances or point process convergence. From a Pareto process perspective, it is convenient to fix a high threshold function $u(s)$ and to assume that
\begin{equation}\label{eq:margdistrib}
{\rm pr}\{X(s)> x\} =   \left[1+ \xi(s)\{x-\mu(s)\}/\sigma(s)\right]_+^{-1/\xi(s)}, \quad x > u(s), 
 \end{equation}
corresponding to the univariate tail probabilities of the generalized Pareto process in~\eqref{eq:trafoppclassgen}, with real parameters $\mu(s)< u(s)$, $\sigma(s)>0$ and $\xi(s)$, such that the right-hand side of~\eqref{eq:margdistrib} is less than unity.

\subsection{Elliptical extremes}\label{sec:ell}

A random vector $X\in\mathbb{R}^D$ is said to follow an elliptical distribution if it can be written as
\begin{equation}
\label{eq:defell}
  X = R A U +  \mu,
\end{equation}
with $R$ a nonnegative random variable, $A$ a $D\times D$ deterministic nonsingular matrix defining the dispersion matrix $\Sigma= A  A'$,  $ U$ a random vector independent of $R$ and distributed uniformly on the Euclidean unit sphere $\{ x\in \mathbb{R}^D :  x'  x =1\}$ and $ \mu\in \mathbb{R}^D$ a deterministic shift vector. The restriction to nonsingular square matrices $A$ excludes some special cases of minor practical importance. Examples of elliptical distributions are the multivariate Gaussian and the multivariate $t$ distributions. As an extension of~\eqref{eq:defell}, a random process $X$ is called elliptical if all its finite-dimensional distributions are elliptical with dispersion matrices $\Sigma$ defined through a covariance function. The max-stable limits in~\eqref{eq:convms} for elliptical processes are either processes with independent univariate marginal distributions in the case of asymptotic independence, as for instance the limits of Gaussian processes, or are extremal-$t$ processes in all other cases. In terms of unit Fr\'echet margins, extremal-$t$ processes can be represented as  
\begin{equation}\label{eq:spectrt}
Z^*(s) =  m_\alpha \max_{i=1,2,\ldots }  W_i(s)_+^\alpha /Q_i, \quad m_\alpha=\pi^{1/2}2^{1-\alpha/2}\Gamma\{(\alpha+1)/2\}^{-1} , 
\end{equation}
where $0<Q_1<Q_2<\cdots$ are the points of a unit-rate Poisson process on the positive half-line, and $W_i=\{W_i(s)\}$ are independent replicates of a standard Gaussian process with continuous sample paths and correlation function $\varrho$ \citep{Opitz:2013}. In particular, $\alpha=1$ yields the extremal Gaussian process \citep{Schlather:2002}. By interpreting the processes $W_i$ as independent marks of the points of  the Poisson process $\{Q_i\}$, we see that the point process $\{P_i\}=\{m_\alpha (W_i)_+^\alpha /Q_i\}$ is Poisson with intensity measure $\Lambda_+$. We use this for simulation from the corresponding $\ell$-Pareto process, see Section~\ref{sec:simu}. We use the term elliptical $\ell$-Pareto process since the tails of its finite-dimensional distributions correspond to elliptical distributions with a Pareto-distributed radial variable $R$ in~\eqref{eq:defell}.
The finite-dimensional dependence structure associated to $D$ sites $ s=(s_1,\ldots ,s_D)$ is characterized by the exponent function \citep{Nikoloulopoulos.etal:2009}
\begin{eqnarray}\label{eq:vell}
V( z) &=& -\log {\rm pr}\{ Z^*(s_1) \leq z_1, \ldots, Z^*(s_D) \leq z_D\}\nonumber \\ 
&=& \sum_{j=1}^D z_j^{-1} t_{\alpha+1}\left\{( z_{-j}/z_j)^{1/\alpha} ; {\Sigma}_{-j,j}, (\alpha+1)^{-1}\left({\Sigma}_{-j,-j}-{\Sigma}_{-j,j} {\Sigma}_{-j,j}'\right)\right\},
\end{eqnarray}
with the correlation matrix $\Sigma=\{\varrho(s_{j_1},s_{j_2})\}_{1\leq j_1,j_2\leq D}$ related to the correlation function $\varrho$, and with $t_\alpha(\cdot;  \mu, \Sigma)$ the cumulative distribution function of a multivariate $t$  distribution with $\alpha$ degrees of freedom and parameters $\mu$ and $\Sigma$.

Dependence structures of Brown--Resnick type arise as a special case of extremal-$t$ dependence when $\alpha\rightarrow\infty$. By analogy with~\eqref{eq:spectrt}, a standard Brown--Resnick process is constructed as $Z_{\text{BR}}^*(s) = \max_{i=1,2,\ldots}\exp\{\tilde{W}_i(s)-\gamma(s)\}/Q_i$ \citep{Kabluchko.etal:2009}, where $\tilde W_i$ are independent and identically distributed copies of an intrinsically stationary centered Gaussian process characterized by its variogram $2\gamma(s)=\E\{\tilde{W}_1(s)^2\}$ and with $\tilde{W}_1(0)=0$ almost surely. For processes $W_1$ whose correlation function $\varrho_\alpha$ depends on $\alpha$ such that the limit $\gamma(s_2-s_1) = \lim_{\alpha\rightarrow\infty} \alpha \{1 - \varrho_\alpha (s_1,s_2)\}$ exists and satisfies $0<\gamma(s_2-s_1)<\infty$ for all sites $s_1,s_2$ with $s_1\not=s_2$, the extremal-$t$ process $Z^*$ in~\eqref{eq:spectrt} converges to $Z_\text{BR}^*$ as $\alpha\rightarrow\infty$ \citep{Nikoloulopoulos.etal:2009}.  For instance, the correlation function $\varrho_\alpha (s_1,s_2)=\exp[-\{\|s_1-s_2\| /(\alpha^{1/\kappa}\lambda)\}^{\kappa}]$ with $\kappa\in(0,2]$ and $\lambda>0$ yields the variogram $2(\|s_1-s_2\|/\lambda)^\kappa$.

The truncation of $W_i$ at zero in~\eqref{eq:spectrt} implies that the measure $\Lambda_+$ of an extremal-$t$ process has positive mass on the set $\{f\in C_+(K)\setminus\{0\}:\min_{s\in K}f(s)=0\}$, which is not the case for Brown--Resnick processes. We later discuss the implications for inference on elliptical $\ell$-Pareto processes.

\section{Inference}\label{sec:inference}

\subsection{Likelihoods for $\ell$-Pareto processes}\label{sec:estim}

We now consider a collection $ s=(s_1,\ldots,s_D)$ of sites in $K$, and $X_{s,1},\ldots,X_{s,n}$ independent replicates of a finite-dimensional observation vector $X_{s}=\{X(s_1),\ldots,X(s_D)\}$, which is embedded in a process $X$. We suppose that $X$ is in the maximum domain of attraction of a max-stable process $Z$. We assume the marginal parameters $\mu(s)$, $\sigma(s)$ and $\xi(s)$ in~\eqref{eq:margdistrib} were estimated in a first step, and we consider the standardized process $X^*$, whose finite-dimensional vectors relative to $s$ are denoted by $X^*_{ s}$. Here we describe the estimation of $\Lambda_+$ based on $\ell$-exceedances of $X^*_{ s}$ with a suitably chosen risk functional $\ell$.  We use the elliptical Pareto process with a parametric correlation function defining a parametric model for the measure $\Lambda_+$.

Different choices of $\ell$ yield different approaches to inference, but it is crucial that $\ell(X^*)$ can be determined from $X_s^*$, so we need $\ell(X^*)=\ell(X_s^*)$. Without loss of generality, we define the exceedance observations in terms of $\ell(X_s^*)\geq 1$. We approximate the distribution of the points $X_s^*$ with $\ell(X_s^*)\geq 1$ by the distribution of the elliptical $\ell$-Pareto process $Y_\ell^*$. 

For a standard $\ell$-Pareto process $Y_\ell^*$, the density of the vector $Y_{s}^*=\{Y_\ell^*(s_1),\ldots,Y_\ell^*(s_D)\}$ on $\{y\in\mathbb{R}_+^D\setminus\{0\}:\ell(y)\geq 1\}$ is $\lambda_{+,s}(y)/\kappa_\ell(K)$,  where $\lambda_{+,s}$ is the density of $\Lambda_{+,s}$, the finite-dimensional marginal measure of $\Lambda_+$ relative to the sites $s$.  When $\Lambda_{+,s}$ is absolutely continuous with respect to Lebesgue measure, $\lambda_{+,s}$ is the full derivative $-V_{1:D}(y)$ of the negated exponent function $V$. Otherwise, when $\Lambda_{+,s}$ puts positive mass on lower-dimensional subspaces of $\mathbb{R}_+^D$, we get slightly different expressions for $\lambda_{+,s}$ on those subspaces \citep{Coles.Tawn:1991}. In the extremal elliptical model, we find positive mass on $\{y \in \mathbb{R}_+^D\setminus \{0\}: \|y\|_\infty > 0\}$, see~\S\ref{sec:formulae}. 

Based on the sample of $\ell$-exceedances $X_{s,k}^*$ $(k=1,\ldots,N_u)$ satisfying $\ell(X_{s,k}^*)\geq 1$, and assuming a parametric $\ell$-Pareto model with parameter vector $\psi$, we obtain the full likelihood
\begin{equation}\label{eq:likgeneral}
\tilde L_\ell(\psi) = \prod_{k=1}^{N_u} \dfrac{\lambda_{+,s}(X_{ s,k}^*)}{\kappa_\ell(K)}.
\end{equation}
When $\kappa_\ell(K)$ cannot be calculated explicitly, Monte Carlo approximations are required to evaluate the likelihood function~\eqref{eq:likgeneral}.
For a choice of $\ell$ that is both tractable and useful in practice, we focus on $\ell(f)=\max_{j=1,\ldots,D} f(s_j)/u_j$ with a high multivariate threshold $u=(u_1,\ldots,u_D)>0$, which select observations for which at least one component exceed its marginal threshold. Then, $\kappa_\ell(K)=V(u)$ and \citep{Ferreira.deHaan:2012}
\begin{equation*}\label{eq:probaexc}
{\rm pr}(  Y_{ s}^* \leq  y ) = \dfrac{V\{\min(y,u)\}-V( y)}{V( u)},\quad  y \not\leq  u,
\end{equation*}
which is the multivariate Pareto distribution defined by \citet{Rootzen.Tajvidi:2006}. Specifying $\kappa_\ell(K)=V(u)$ in \eqref{eq:likgeneral} yields the corresponding likelihood 
\begin{equation*}\label{eq:unlik}
\tilde L_1(\psi) = \prod_{k=1}^{N_u} \dfrac{\lambda_{+,s}(X_{ s,k}^*)}{V(u)}.
\end{equation*}
Inference based on $\tilde L_1$ might be compromised in practice: first, using the full information from an observation $X^*_{s,k}$ with $\ell( X^*_{s,k})\geq 1$ might be inefficient since the asymptotic distribution might model  the non-exceeding components badly and thus induce bias in the estimators. Second, positive mass on the boundary of $\mathbb{R}_+^D\setminus \{0\}$ creates a discontinuity due to the weak convergence of the data process to the $\ell$-Pareto process in \eqref{eq:convellPareto}, as is the case for the elliptical model. The margins of $X^*_{s,k}$ are standard Pareto, and so are strictly positive, which is incoherent with the possible mass on the axis for $\Lambda_{+,s}$. To overcome these two issues, we propose the use of a censoring scheme. We consider the censored observations $X^c_{s,k}=\max(X_{s,k}^*,u)$, where the maximum is taken componentwise. The {corresponding likelihood is
\begin{equation*}\label{eq:Paretolik}
\tilde L_2(\psi) =  \prod_{k=1}^{N_u} \dfrac{-V_{I_k}(X_{s,k}^c)}{V(u)},
\end{equation*}
where $V_{I_k}$ denotes the partial derivative of $V$ with respect to the indices  $I_k\subset\{1,\ldots,D\}$ associated to the components that exceed their corresponding marginal thresholds. 

When both $n$ and $N_u$ are observed, we propose to incorporate the information provided by the binomial variable $n-N_u$, that represents the number of fully-censored observations. We use the approximation $\pr\{\ell(X^*)\geq 1\}=\pr\{ \max_{j=1,\ldots,D} X^*(s_j)/u_j \geq 1\} \approx V(u)$, which follows from~\eqref{eq:convms} for a high threshold vector $u$ \citep[][Theorem~9.3.1]{deHaan.Ferreira:2006}, and define the likelihoods
\begin{equation*}
L_m= \{1-V( u)\}^{n-N_u}V(u)^{N_u} \times \tilde L_m \quad (m=1,2).
\end{equation*}
The threshold vector $u$ must be high enough to yield $V( u)\leq 1$.

Full likelihood inference based on $L_1$ or $L_2$ is possible if  $\lambda_{+,s}$, the function $V$ and its partial derivatives, are known. We derive these expressions for elliptical Pareto processes in Section~\ref{sec:formulae}; expressions for Brown--Resnick processes were derived by \citet{Wadsworth.Tawn:2013}. By contrast, inference for max-stable extremal-$t$ and Brown--Resnick processes is typically based on composite likelihoods \citep{Padoan.etal:2010}. In Section~\ref{sec:eff}, we use simulation to investigate the gain in efficiency from the use of full likelihoods. 

We relate our approach to \citet{Wadsworth.Tawn:2013} and \citet{Engelke.etal:2012}, who proposed full likelihood inference based on the finite-dimensional convergence to the Poisson process in~\eqref{eq:convPP}. \citet{Wadsworth.Tawn:2013} proposed a censored approach with the likelihood
\begin{equation*}
\exp\{-nV(u)\} V(u)^{N_u}  \times \tilde L_2(\psi), 
\end{equation*}
which differs from $L_2$ only through the distribution assumed for the number of exceedance $N_u$: binomial for the Pareto approach, and Poisson for the point process approach. Since $n$ is large and $V(u)$ is small in practice, these two approaches give very similar results. By contrast, \citet{Engelke.etal:2012} considered $X_{s,k}^*$ as an exceedance when $\sum_{j=1}^D X_{s_j,k}^*>u$, leading to inference based on the multivariate sum spectral measure \citep{Coles.Tawn:1991}. An equivalent approach in the framework of Pareto processes is obtained for $\ell(f)= \sum_{j=1}^D f(s_j)/u$ for a threshold $u>0$, where $\kappa_\ell(K)=D/u$ in the likelihood~\eqref{eq:likgeneral}. Although this approach seems to perform well for Brown--Resnick processes, it would be inefficient for elliptical Pareto processes due to the singularities in $\Lambda_{+,s}$, just like the uncensored likelihood $L_1$; see Section~\ref{sec:eff}. \citet{Engelke.etal:2012} further considered the use of extremal increments, corresponding to $\ell(f)=f(s_0)/u_0$ for fixed $s_0\in K$ and threshold $u_0>0$, but this approach has the same disadvantages in the case of elliptical Pareto processes. 

\subsection{Densities and partial derivatives of the exponent function for extremal-$t$ processes}\label{sec:formulae}
We derive the density $\lambda_{+,s}$ of the finite-dimensional exponent measure $\Lambda_{+,s}$ and the partial derivatives $V_{I_k}$ through calculations similar to those of \citet{Wadsworth.Tawn:2013} for Brown--Resnick processes. A complication for extremal-$t$ processes arises from the singularities of $\Lambda_{+,s}$ on the boundary of $\mathbb{R}_+^D\setminus \{0\}$. 
To resolve this, we observe that the extremal-$t$ process $Z^*$ in~\eqref{eq:spectrt} arises as the pointwise maximum of a Poisson process with points $\tilde P_i=m_\alpha T_\alpha(W_i)/Q_i$, where $T_a( x) = {\rm sign}(x) | x |^a$ for $a>0$. The truncation of $W_i$ at zero in~\eqref{eq:spectrt} is irrelevant because $Z^*$ is constituted from pointwise maxima that are positive almost surely. If the point process of the $\tilde P_i$ has intensity measure $\Lambda$, the unique measure $\Lambda_+$ of the extremal-$t$ process is obtained by projecting the negative values to zero. Therefore, we first calculate the intensity $\lambda_{s}(y)$ for $y\in\mathbb{R}^D\setminus\{0\}$. To derive $\lambda_{+,s}(y)$ when some components of $y$ are zero, say $y=(\tilde{y},0)$ with $\tilde{y}>0$, we can integrate $\lambda_{s}$ over all negative values of the zero-components in $y$ such that $\lambda_{+,s}(y) = \int_{-\infty}^{0} \lambda_{s}(\tilde{y}, z)\rm d  z$.

\citet{Ribatet:2013} gives the density $\lambda_s$ of $\Lambda_{s}$,
\begin{align*}
\lambda_{ s}( y)&=\alpha^{1-D}\pi^{(1-D)/2}| \Sigma_{ s}|^{-1/2} \Gamma\{(\alpha+1)/2\}^{-1}\Gamma\{(\alpha+D)/2\} \\
&\quad \times \prod_{j=1}^D |y_j|^{1/\alpha-1} \{T_{1/\alpha}( y)' \Sigma_{ s}^{-1} T_{1/\alpha}( y)\}^{-(\alpha+D)/2}, \quad y\in\mathbb{R}^D,
\end{align*}
where $\Sigma_{ s} = \{\varrho(s_{j_1},s_{j_2})\}_{1\leq j_1,j_2\leq D}$ denotes the finite-dimensional correlation matrix stemming from the correlation function $\varrho$ of the extremal-$t$ dependence structure relative to the sites $s=(s_1,\ldots,s_D)$. The density $\lambda_{+,s}$ of $\Lambda_{+,s}$ on $(0,\infty)^D$ equals $\lambda_{ s}$.
The partial derivatives $V_{I_k}$ of the exponent function $V$ are calculated by integrating $\lambda_{ s}$ with respect to the components in the set complementary to $I_k$. The integration is carried out using conditional intensities. Given a collection $ s_0=(s_{0,1},\ldots,s_{0,d})$ of $d$ conditioning locations with values $y_0$, the conditional intensity $\lambda_{ s\mid  s_0, y_0}( y)=\lambda_{( s, s_0)}( y, y_0)/\lambda_{ s_0}( y_0)$ equals \citep{Ribatet:2013}
\begin{align}\label{eq:conddensity}
\lambda_{ s\mid  s_0, y_0}( y) &= \alpha^{-D} \pi^{-D/2} (d+\alpha)^{-D/2} |{\tilde{\Sigma}}|^{-1/2} \Gamma\{(\alpha+d)/2\}^{-1} \Gamma\{(\alpha+D+d)/2\} \nonumber \\ 
&\quad \times \prod_{j=1}^D |y_j|^{1/\alpha-1} \left[1+\dfrac{\{T_{1/\alpha}( y)-{\tilde\mu}\}' {\tilde\Sigma}^{-1} \{T_{1/\alpha}( y)-{\tilde\mu}\}}{d+\alpha}\right]^{-(\alpha+D+d)/2},
\end{align}
with
\begin{equation*}
{\tilde\mu}=\Sigma_{ s: s_0} \Sigma_{ s_0}^{-1} T_{1/\alpha}( y_0),\quad {\tilde\Sigma}= \dfrac{ T_{1/\alpha}(y_0)' \Sigma_{ s_0}^{-1}  T_{1/\alpha}(y_0)}{d+\alpha} \left(\Sigma_{ s}-\Sigma_{ s: s_0} \Sigma_{ s_0}^{-1} \Sigma_{ s_0: s}\right),
\end{equation*}
where $\Sigma_{ s: s_0}$ denotes the matrix of covariances between the random vectors corresponding to the location vectors $s$ and $s_0$. Expression~\eqref{eq:conddensity} is the density of a random vector $T_\alpha( X)$, where $ X$ follows a $D$-dimensional $t$ distribution with $d+\alpha$ degrees of freedom and parameters ${\tilde\mu}$ and ${\tilde\Sigma}$.

Without loss of generality,  we consider the partial derivative $V_{1:d}( y)$ of $V$ with respect to the indices $1$ to $d$ such that $I_k=\{1,...,d\}$, obtained by calculating the integral of  $\lambda_{s_{(d+1):D} \mid s_{1:d}, y_{1:d}}( y_{(d+1):D})$ and by multiplying the resulting expression by $\lambda_{s_{1:d}}( y_{1:d})$. The required integral of the conditional density is $t_{d+\alpha}( y_{(d+1):D}^{1/\alpha};\tilde\mu,{\tilde\Sigma})$. We get 
\begin{align}\label{eq:partialV}
-V_{1:d}( y)=& t_{d+\alpha}\left( y_{(d+1):D}^{1/\alpha};\tilde\mu,\tilde\Sigma\right) \alpha^{1-d}\pi^{(1-d)/2}|\Sigma_{1:d}|^{-1/2} \Gamma\{(\alpha+1)/2\}^{-1} \nonumber \\
&\times \Gamma\{(\alpha+d)/2\}\left(\prod_{j=1}^d |y_j|\right)^{1/\alpha-1} \big\{ (y_{1:d}')^{1/\alpha} \Sigma_{1:d}^{-1}  y_{1:d}^{1/\alpha}\big\}^{-(\alpha+d)/2},
\end{align}
with ${\tilde\mu}=\Sigma_{(d+1):D,1:d} \Sigma_{1:d}^{-1}  y_{1:d}^{1/\alpha}$ and ${\tilde\Sigma}=(d+\alpha)^{-1}  ( y_{1:d}')^{1/\alpha} \Sigma_{1:d}^{-1}  y_{1:d}^{1/\alpha} (\Sigma_{(d+1):D}-\Sigma_{(d+1):D,1:d} \Sigma_{1:d}^{-1} \Sigma_{1:d,(d+1):D})$. Equation~\eqref{eq:partialV} also gives the densities $\lambda_{+,s}$ for  a point $y$ on the boundary of $\mathbb{R}_+^D\setminus\{0\}$: if $y_{1:d}>0$ and $y_{(d+1):D}=0$, then the  density on the corresponding subset of $\mathbb{R}_+^D\setminus\{0\}$ is $-V_{1:d}( y)$, see \citet[][\S3.1]{Coles.Tawn:1991}.

\subsection{Maximum likelihood inference}

Numerical maximization of $L_1$ or $L_2$ yields the maximum likelihood estimate $\hat\psi$ for the vector of  parameters $\psi$ of an elliptical Pareto process. Assuming that the data come from the limiting model, standard regularity conditions ensure consistency and asymptotic normality of $\hat\psi$, with an asymptotic covariance matrix that equals the inverse Fisher information matrix \citep[][Ch.~5]{vanderVaart:2000}. In practice, the asymptotic covariance matrix can be estimated by the Hessian matrix of the negated log-likelihood evaluated at $\hat\psi$.

A practical inconvenience for maximum likelihood inference based on $L_1$ or $L_2$ is the need to calculate the $t$ probabilities in \eqref{eq:vell} and \eqref{eq:partialV}. They can be calculated using Monte Carlo approximations \citep{Genz.Bretz:2009}, but the use of full likelihood inference might be too slow if $D>50$. In larger dimensions, one could partition the sample sites into moderately large groups and use a composite likelihood based on the full likelihood contribution from each group.

\section{Exact simulation procedures}\label{sec:simu}
We now describe exact finite-dimensional simulation procedures for extremal-$t$ and elliptical $\ell$-Pareto processes. Due to the elliptical structure of the points $P_i^{1/\alpha}$ from the point process $\{P_i\}$ in~\eqref{eq:spectrt}, an equivalent representation of the finite-dimensional projection of an extremal-$t$ process relative to $D$ sites $s=(s_1,\ldots,s_D)$ is obtained by setting   
\begin{equation}
\label{eq:repmsmv}
P_{s, i}=\{\E(U_{1,1})_+^\alpha\}^{-1} ( A_{ s} {U_i})_+^\alpha/Q_i,
\end{equation}
with $A_{ s}$ the Cholesky root of the correlation matrix $\Sigma_{ s}= A_{ s}  A'_{ s}$ and $U_i=(U_{i,1},\ldots,U_{i,D})'$ independent and identically distributed copies of a vector $ U$ uniformly distributed on the Euclidean unit sphere \citep[][Theorem~3.2]{Opitz:2013}. This allows exact simulation of both max-stable and Pareto processes due to the boundedness $\|( A_{ s} {U_i})_+^\alpha\|_\infty \leq 1$.

In practice, max-stable processes are simulated using only a finite number of $ P_{ s,i}$; see Fig.~\ref{fig:FigSimMSPareto}. When  a finite boundary $b< \infty$ exists for the components of $Q_i  P_{ s,i}$ such that ${\rm pr}\{ \max_{i=1,2,...}Q_i P_{ s,i}(s_j) \leq b\} = 1$ $(j=1,...,D)$, exact simulation of $ Z_{ s}^*$ can be achieved from a finite number of points $ P_{ s,i}$ \citep[][Theorem 4]{Schlather:2002}. Since the components of $\{\E(U_{1,1})_+^\alpha\}^{-1} ( A_{ s} {U_i})_+^\alpha$ in \eqref{eq:repmsmv} are always bounded by $b=\{\E(U_{1,1})_+^\alpha\}^{-1}$, exact simulation of extremal-$t$ processes is possible. For $i\geq 1$, $\| P_{ s, i} \|_\infty \leq b/Q_{i}$ with an increasing sequence $\{Q_i\}$.  If $\| \max_{i=1,\ldots,\tau_b}P_{s, i}\|_\infty \geq b/Q_{\tau_b}$ for some $\tau_b>1$, then the points $P_{s, i}$ for $i>\tau_b$ cannot contribute to the maximum in~\eqref{eq:spectrt} and we have $Z_s^*=\max_{i=1,\ldots,\tau_b}P_{s, i}$.  Two numerical limitations may restrict  the applicability of this simulation approach: first, standard algorithms for determining the Cholesky root $ A_{ s}$ of $ \Sigma_{ s}$ require $O(D^3)$ basic operations; second, $b$ may be large if $\alpha$ or $D$ are large, requiring the simulation of  a very large number of points $ P_{ s,i}$. 
More precisely, 
\begin{equation*}
 b = 2\pi^{1/2} \dfrac{\Gamma\{(D+\alpha)/2\}}{ \Gamma\{(\alpha+1)/2\}\Gamma(D/2)} \approx 2^{1-\alpha/2} \pi^{1/2} \dfrac{(D+\alpha-2)^{\alpha/2}}{\Gamma\{(\alpha+1)/2\}},\quad D\rightarrow\infty,
\end{equation*}
using Stirling's formula. In certain situations, notably when $D$ indexes a fine spatial grid of points, these limitations are too restrictive. Then the conventional approach for approximate simulation can be used. Since the tails of $W(s_j)_+^\alpha$ become heavier when $\alpha$ increases, the approximation error in the simulated max-stable process also increases. 

The simulation of the points $P_{ s,i}$ in~\eqref{eq:repmsmv} yields an algorithm for the simulation of elliptical $\ell$-Pareto processes: as mentioned in~\S\ref{sec:ell}, the points $P_{ s,i}$ with $\ell(P_{ s,i})\geq 1$ are independent realisations from the standard $\ell$-Pareto process; see Fig.~\ref{fig:FigSimMSPareto}. Moreover, for $u_0>0$, the homogeneity of $\Lambda_{+,s}$ implies that the points $u_0^{-1} P_{ s,i}$ with $\ell(P_{ s,i})\geq u_0$ are also realisations from the standard $\ell$-standard process. The existence of the upper bound $b$ allows us to simulate all the points $P_{ s,i}$ in a set $A=([0,u]^D)^C\subset\mathbb{R}_+^D$ for $u>0$. Since the set  $\ell(y)\geq u_0$ is a subset of $A$ for suitably chosen $u$, we can obtain exact simulations from every elliptical $\ell$-Pareto process.

\begin{figure}[t]
\centerline{\includegraphics{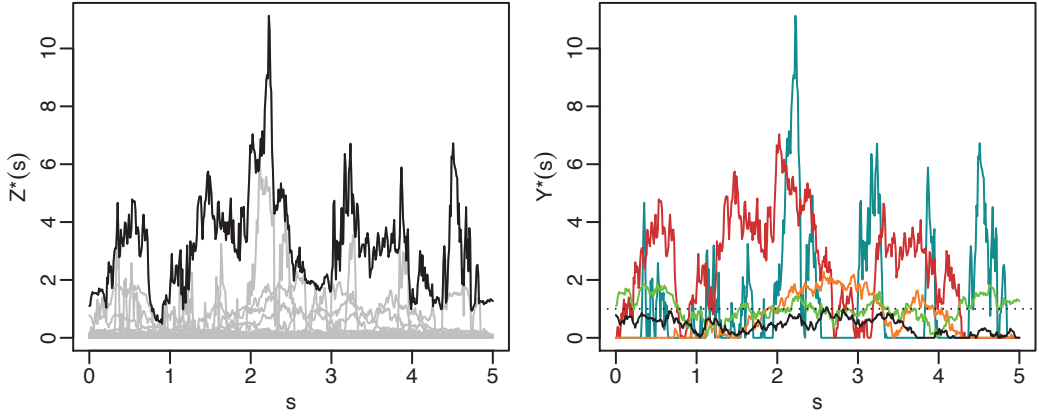}}
\caption{Left: simulation (black line) from an extremal-t process with $\alpha=1$ and $\varrho(h)=\exp(-\|h\|)$. The grey lines show the points $P_{s,i}$ in the spectral decomposition~\eqref{eq:repmsmv}. Right:  independent simulations from the corresponding elliptical $\ell$-Pareto process with $\ell(f)=\sup_{s\in[0,5]}f(s)$ are given by the points $P_{s,i}$ with $\ell(P_{s,i})\geq 1$.}
\label{fig:FigSimMSPareto}
\end{figure}

Instead of simulating the points of the Poisson process, it is possible to use an acceptance-rejection algorithm to generate realisations of $Y^*_\ell(s)$ without  dealing with a random number of realisations. 
First, we consider the simulation of a standard $\ell$-Pareto process with $\ell(f)=\max_{j=1,...,D} f(s_j)$. From the previous paragraph, the points $b^{-1}P_{ s,i}$ with $\ell(P_{ s,i})\geq b$ are realisations from the standard $\ell$-Pareto process.
The condition $\ell(P_{s, i})\geq b$ implies $1/Q_i\geq 1$, hence only the points $P_{s, i}$ with $1/Q_i\geq 1$ need to be simulated. Since the points $1/Q_i$ satisfying $1/Q_i\geq 1$ are distributed according to the standard Pareto distribution, any vector $R (A_s U)_+^\alpha$ with  $U$ independent of a standard Pareto distributed random variable $R$ is a standard $\ell$-Pareto process if  $\ell\{R(A_s U)_+^\alpha\}\geq 1$. 
When $\ell$ is different from the componentwise maximum, we proceed as before and fix $u_0>0$ such that $\max_{j=1,...,D} f(s_j)\geq 1$ whenever $\ell( f)\geq u_0$. Then  the vector $u_0^{-1}R (A_s U)_+^\alpha$, given that $\ell\{R (A_s U)_+^\alpha\}\geq u_0$, is a realisation of $Y^*_\ell(s)$. We can get a sample  of $Y^*_\ell(s)$ by repeatedly simulating random vectors $u_0^{-1}R (A_s U)_+^\alpha$ and retaining only those vectors fulfilling the condition  $\ell\{R (A_s U)_+^\alpha\}\geq u_0$. To minimise the frequency of rejections, $u_0$ should be chosen as small as possible.

Whereas conditioning $\Lambda_+$  on exceedances of  $\ell( f)$ over unity yields  the distribution of the $\ell$-Pareto process, one might instead be interested in the conditional distribution when values $ y_0> 0$ for a collection of sites $ s_0=(s_{0,1},\ldots,s_{0,d})$ are fixed. The finite-dimensional conditional distribution for the sites $ s =(s_1,\ldots,s_D)$, disjoint with $s_0$, has density \eqref{eq:conddensity}. The conditional process defined on $K\setminus\{s_0\}$ corresponds to a transformed $t$ process that can easily be simulated. 

\section{Simulation study}\label{sec:eff}

We used simulation to investigate the efficiency of the full likelihoods $L_1$ and $L_2$ for estimating the parameters of elliptical Pareto processes. For comparison, we also report results from the pairwise censored likelihood approach based on all pairs, which represents a standard approach to fitting max-stable models. Based on the exact simulation procedure introduced in Section~\ref{sec:simu}, we generated samples of $1000$ elliptical $\ell$-Pareto processes with $\ell( f)=\max_{s\in K} f(s)$ at $16$ locations given by $K=\{(s_1/3,s_2/3)\}_{s_1,s_2 \in \{0,1,2,3\}}$. We chose the stable correlation function $\varrho(h)=\exp\{-(\|h\|/\lambda)^\kappa\}$, where $h$ is the lag vector between two locations, $\lambda>0$ is a range parameter and $0<\kappa\leq 2$ is a smoothness parameter. Different combinations of values for the degrees of freedom $\alpha>0$ and for $\lambda$ and $\kappa$ were considered,  covering small to strong dependence with different degrees of smoothness. We estimated the vector of parameters $\psi=(\log\lambda,\kappa,\alpha)$ using the three approaches, each of them based on marginal thresholds equal to the $95\%$ quantiles. The mean squared error of $\hat\psi$ can be decomposed into a sum of bias and variance terms: $\mathrm{MSE}(\hat\psi)=\|\E(\hat\psi)-\psi\|^2 + \mathrm{tr}(V)$, where $V$ is the covariance matrix of $\hat\psi$. For each parameter configuration, estimates for $1000$ samples were calculated to obtain the bias and covariance matrix of each estimator. Table~\ref{tab:eff-pareto} shows the relative efficiency of the three estimators, here defined as the ratio of the trace of their covariance matrices. Unreported results showed that all estimators have only little or no bias. Throughout, the full uncensored likelihood estimator $L_1$ was found to be more efficient than the full censored estimator $L_2$, owing to the loss of information from censoring. The difference is larger when dependence is weak, that is, when more components are censored in exceedances, though more exceedances are observed. The $L_2$-based estimator is more efficient than its pairwise equivalent, and efficiency gains are larger for smooth processes with weak dependence. Overall, the relative reduction in variance is around $60\%$. Other simulations indicated that the efficiency improvements of the full likelihood over the pairwise likelihood become larger when the number of sites is larger: in a similar estimation framework, we found a reduction of variance of around $35\%$ for nine locations and of around $10\%$ for four locations.

\begin{table}
\caption{Relative efficiency (in \%) of full and pairwise likelihood estimators for the parameters of elliptical Pareto processes with the stable correlation function. For each combination of $\kappa$ and $\alpha$, three values of $\lambda$ were chosen to give pairwise extremal coefficients $\theta\in\{1$$\cdot$$2,1$$\cdot$$4,1$$\cdot$$6\}$ at distance $0$$\cdot$$5$. Each cell gives the ratio of the covariance matrix traces for the uncensored and censored full likelihood estimators and for the censored full and pairwise  likelihood estimators, separated by $/$.}
\vspace*{-0.3cm}
\begin{center}
\footnotesize
\begin{tabular}{l cccc}  
& \multicolumn{4}{c}{$\kappa= 0.5 $} \\ 
$\theta/\alpha$ & $1$ & $2$ & $5$ & $10$\\
$1.2$ & $61/62$ & $51/59$ & $48/60$ & $45/58$ \\ 
$1.4$ & $48/55$ & $29/51$ & $16/52$ & $15/53$ \\ 
$1.6$ & $30/50$ & $14/39$ & $ 5/45$ & $ 4/39$ \\ 
\end{tabular} 
\,
\begin{tabular}{cccc}  
\multicolumn{4}{c}{$\kappa= 1 $} \\ 
$1$ & $2$ & $5$ & $10$\\
$50 /45$ & $39/47$ & $31/42$ & $36/41$ \\ 
$41 /43$ & $22/37$ & $ 8/35 $ & $ 7/44$ \\ 
$34 /36$ & $13/35$ & $ 3/31$ & $ 2/37$ \\ 
\end{tabular} 
\,
\begin{tabular}{cccc}  
\multicolumn{4}{c}{$\kappa= 1.5 $} \\ 
$1$ & $2$ & $5$ & $10$\\
$43/29$ & $27/28$ & $21/27$ & $27/21$ \\ 
$34/27$ & $18/23$ & $ 5/19$ & $ 5/21$ \\ 
$34/25$ & $16/22$ & $ 3/19$ & $ 1/25$ \\ 
\end{tabular} 
\end{center}
\label{tab:eff-pareto}
\end{table}

To investigate the impact of the convergence to a limiting elliptical Pareto process, we further simulated samples $X_{s,1},\ldots,X_{s,1000}$ of $t$ processes with $\alpha$ degrees of freedom on the same grid as before. Marginal distributions of $X_{s,k}$ were transformed to the standard Pareto scale by the transformation $X_{s,k}^*=1/\{1-t_\alpha(X_{s,k})\}$ ($k=1,\ldots,1000$), where $t_\alpha$ denotes the cumulative distribution function of a univariate $t$ variable with $\alpha$ degrees of freedom}. We fitted elliptical $\ell$-Pareto processes to threshold exceedances of the simulated $X_{s,k}^*$ over the marginal $95\%$, $98\%$ and $99\%$ quantiles using the two full likelihoods $L_1$ and $L_2$ and the pairwise censored likelihood. We then considered the bias, variance and mean squared error of these estimators. As opposed to the simulations discussed in the previous paragraph, the elliptical Pareto model is only valid asymptotically and so the estimators are biased. Table~\ref{tab:eff-t} reports the bias and the empirical covariance matrix trace of $\hat\psi$ calculated from $1000$ estimates and the relative efficiencies of the two full likelihood and the two censored likelihood estimators, here defined as the ratio of their mean squared error. For all thresholds, the uncensored estimator $L_1$ has the largest mean squared error because of its very large bias. The two censored estimators have small bias when $\alpha=1$, but the bias increases as $\alpha$ increases. This may be explained by the slower convergence to the limiting dependence structure for larger $\alpha$. The bias is reduced by increasing the thresholds such that the exceedance distribution is closer to the asymptotic model; variances increase accordingly. Variances are always smaller for the full likelihood estimator than for the pairwise one, but the bias of the full likelihood estimator is often larger. In terms of mean squared error for the $95\%$ threshold, the full likelihood estimator outperforms the pairwise one for $\alpha<6$, but not otherwise. The bias of the full likelihood estimator decreases for higher thresholds, and the full likelihood estimator generally has a smaller mean squared error than the pairwise estimator owing to its smaller variance. Hence for large values of $\alpha$, very high thresholds are needed for the full likelihood estimator to outperform the pairwise estimator in terms of mean squared error. 

The results of these simulations suggest that censored approaches are the best in practice when the model is misspecified. Moreover, full likelihood inference improves estimation efficiency when the distribution of extremes is close to the limiting model, but the pairwise approach appears more robust to certain kinds of model misspecification.

\begin{table}[t]
\caption{Estimation of elliptical $\ell$-Pareto processes based on exceedances of $t$ processes for the stable correlation function with $\kappa=1$. For each $\alpha$ in $1,\ldots,10$, the values of $\lambda$ were chosen to yield the pairwise extremal coefficient $\theta=1$$\cdot$$4$ at distance $0$$\cdot$$5$. For each of the thresholds chosen at the $95\%$, $98\%$ and $99\%$ quantiles and for each of the three estimators based on $L_1$, $L_2$ or the pairwise censored distributions, the bias/variance terms of $\hat\psi$ are reported. For each threshold, the last row reports the ratio $\delta$ of the mean squared error for the censored and uncensored full likelihood estimators and for the censored full and pairwise likelihood estimators, separated by $/$. All numbers have been multiplied by $100$.}
\vspace*{-0.3cm}
\begin{center}
\footnotesize
\begin{tabular}{ll cccccccccc}  
& $\alpha$ & $1$ & $2$ & $3$ & $4$ & $5$ & $6$ & $7$ & $8$ & $9$ & $10$ \\ 
$ 95 \% $& $L_1$ & $303/  9$ & $241/  6$ & $223/  5$ & $215/  6$ & $210/  7$ & $208/  8$ & $205/ 10$ & $203/ 10$ & $202/ 10$ & $201/ 11$\\
& $L_2$ & $2/3$ & $12/ 6$ & $34/ 9$ & $53/11$ & $69/14$ & $82/16$ & $92/21$ & $100/ 21$ & $108/ 23$ & $115/ 25$\\
& pw & $3/7$ & $15/14$ & $28/20$ & $43/27$ & $55/32$ & $67/37$ & $71/42$ & $77/44$ & $79/51$ & $82/48$\\
& $\delta$ & $ 0/44$ & $ 1/45$ & $ 4/72$ & $ 8/87$ & $14/98$ & $ 19/101$ & $ 24/114$ & $ 29/117$ & $ 34/124$ & $ 38/135$\\
$ 98 \% $& $L_1$ & $257/ 16$ & $187/ 10$ & $171/  8$ & $170/  9$ & $172/ 12$ & $173/ 13$ & $176/ 18$ & $179/ 20$ & $182/ 24$ & $186/ 26$\\
& $L_2$ & $7/7$ & $ 2/13$ & $13/22$ & $30/28$ & $41/34$ & $57/44$ & $68/53$ & $79/59$ & $90/70$ & $91/70$\\
& pw & $ 6/18$ & $16/38$ & $26/55$ & $42/70$ & $55/86$ & $64/98$ & $ 66/107$ & $ 71/103$ & $ 76/114$ & $ 76/108$\\
& $\delta$ & $ 1/42$ & $ 4/32$ & $ 8/38$ & $13/42$ & $17/44$ & $25/55$ & $30/66$ & $35/78$ & $42/88$ & $41/93$\\
$ 99 \% $& $L_1$ & $240/ 25$ & $160/ 14$ & $141/ 12$ & $142/ 14$ & $145/ 18$ & $151/ 25$ & $157/ 30$ & $162/ 33$ & $169/ 43$ & $174/ 47$\\
& $L_2$ & $12/15$ & $14/26$ & $ 6/40$ & $15/57$ & $23/75$ & $43/97$ & $ 50/114$ & $ 62/115$ & $ 79/171$ & $ 87/205$\\
& pw & $ 8/39$ & $19/70$ & $ 30/111$ & $ 48/140$ & $ 54/175$ & $ 72/208$ & $ 75/234$ & $ 80/220$ & $ 84/277$ & $ 80/237$\\
& $\delta$ & $ 3/41$ & $10/38$ & $19/33$ & $27/36$ & $35/39$ & $46/44$ & $50/48$ & $52/54$ & $71/67$ & $80/93$\\
\end{tabular}
\end{center}
\label{tab:eff-t}
\end{table}

\section{Application}\label{sec:appli}

\begin{figure}[t]
\centerline{\includegraphics{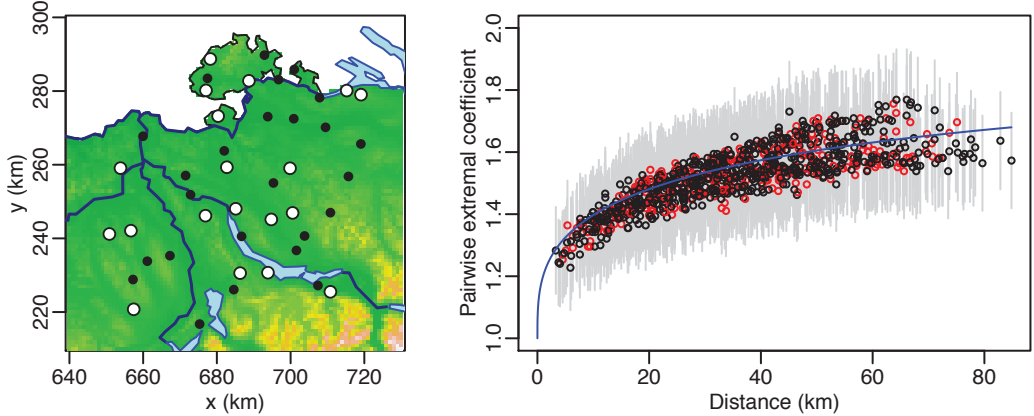}}
\caption{Modelling extreme precipitation around Z\"urich. Left: region of Z\"urich. Black dots correspond to the stations used for the fit, white dots to the stations used for validation. Right: empirical pairwise extremal coefficients (with $95\%$ confidence intervals in grey) for the data used to fit the model (in red) and for the validation data (in black), with the fitted extremal coefficient curve for the best elliptical Pareto model in blue.}
\label{fig:Fig1}
\end{figure}

We illustrate the use of $\ell$-Pareto processes for modelling precipitation extremes in the region of Z\"urich, Switzerland. Daily cumulative rainfall data at $44$ locations were provided by M\'et\'eoSuisse; see Fig.~\ref{fig:Fig1}. Elevations vary from 327 to 718~m for these stations. 
Our analysis is based on summer data recorded from 1 June to 31 August for the years 1962--2012. A preliminary study showed no signs of non-stationarity in the time series and only weak day-to-day dependence in exceedances over the $95\%$ quantiles, leading us to model the daily data as independent and identically distributed. The data seem coherent with the assumption of asymptotic dependence, which suggests modelling threshold exceedances using Pareto processes; see the Supplementary Material. We selected $25$ stations for the fit of the spatial model, see Fig.~\ref{fig:Fig1}; the other stations are kept for  validation. First, we fitted a spatially varying model for the univariate marginal distributions~\eqref{eq:margdistrib} over marginal thresholds taken to be the $95\%$ percentiles at each of the $25$ stations. We used a Bayesian hierarchical model to capture spatial random effects in $\mu(s)$ and $\sigma(s)$, similar to the latent variable model of \citet{Davison.etal:2012}, see the Supplementary Material. The shape parameter $\xi$ was assumed to be constant over the region. Its estimate and $95$\% credible interval is 0$\cdot$11~(0$\cdot$08,0$\cdot$14) corresponding to heavy-tailed marginal distributions. We then transformed the original data at each location to the standard Pareto scale by using the fitted marginal distributions above the thresholds and the empirical distributions below them. 

In a second step, we used the likelihood $L_2$ to model the dependence in the standardized data with $\ell$-Pareto processes for $\ell(f)=\max_{j=1,\ldots,25} f(s_j)/20$; $20$ is the $95\%$ quantile of a standard Pareto distribution. We fitted an elliptical Pareto process with a stable correlation function $\varrho( h)=\exp\{-(\| h\|/\lambda)^\kappa\}$ and shape $\alpha>0$. For comparison, we also fitted the Pareto model of a Brown--Resnick process with $\gamma(h)=(\| h\|/\lambda)^\kappa$, corresponding to the limiting model when $\alpha\rightarrow\infty$; see~\S\ref{sec:ell}. We used the Akaike information criterion to select the best model: for the elliptical Pareto model, it is $864$ less than that of the Brown--Resnick model. The parameter estimates and standard errors for the elliptical Pareto model are $\hat\lambda=$~520~(73)~km, $\hat\kappa=$~0$\cdot$63~(0$\cdot$02) and $\hat\alpha=$~6$\cdot$3~(0$\cdot$4), which yields a process with realisations that are continuous but not differentiable. These results are consistent with those found by \citet{Davison.etal:2012} who identified an extremal-$t$ model with $\hat\alpha=$~5$\cdot$5~(2$\cdot$1) as the best max-stable model for yearly maxima of daily cumulative rainfall on the same region and found a similar estimate for the smoothness parameter of the correlation function.

We validated the accuracy of our model for modelling spatial extremes using the data from the other $19$ stations. The right panel of Fig.~\ref{fig:Fig1} shows estimates of pairwise extremal coefficients related to these stations. The estimates for validation stations are only slightly more variable. Overall, the extremal dependence for validation data is adequately represented by the model. 

Using the conditional distribution~\eqref{eq:conddensity}, we considered simulation at the $19$ validation stations conditional on the values observed at the other stations when at least one of the $25$ components exceeded its marginal $95\%$ threshold. To compare the observed extreme values at the $19$ validation locations with those predicted by the model, we simulated $30000$ conditional realisations for each day, and we measured the proportion of true values falling between the $2.5\%$ and $97.5\%$ quantiles of the simulations. Over the $19$ locations and the $986$ days with at least one exceedance at the other $25$ locations, approximately $85\%$  of the observed values were in the $95\%$ simulated intervals. We do not attain $95\%$ corresponding to a perfect prediction, which can at least partly be explained by the fact that the simulations used conditional data below the thresholds and also ignored the uncertainty of the estimates of the fitted model; simulations taking these two aspects into account should be more variable and thus have a higher coverage probability.

\begin{figure}
\centerline{\includegraphics{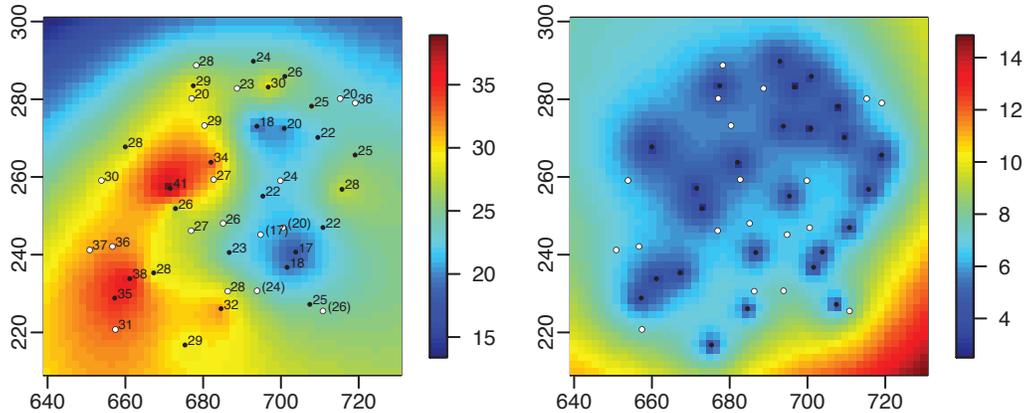}}
\caption{Conditional simulation of extreme rainfall over the region of Z\"urich for 1 June 1962. Left: mean of the conditional simulation of daily cumulative rainfall (in mm). Black dots correspond to the locations on which we conditioned, white dots to the data used for validation with their observed values (written in parentheses under the thresholds). Right: standard deviation for the conditional simulation.}
\label{fig:Fig2}
\end{figure}

Finally, we illustrate the ability of the Pareto process approach to easily simulate conditional rainfall given observed extremes at some set of locations. For a particular day and given the observed data at the $25$ locations used for the fit, we simulated conditional values of rainfall over the region using the transformed $t$ process characterized in~\eqref{eq:conddensity}; the left panel of Fig.~\ref{fig:Fig2} shows the mean of these simulated rainfall fields and the right panel its standard deviation.

\section{Discussion}
Pareto models are appealing because they generalize peaks-over-threshold stability to the spatial context and they appropriately exploit the regularity of the exponent measure in pseudo-polar coordinates for extrapolation on extremes. In this paper we have introduced inference and simulation procedures for the elliptical Pareto model.  Numerical results suggest that the censored approach based on marginal thresholds gives more reliable estimates for this process, but the choice of a sufficiently high threshold is crucial to guarantee that the limiting model provides an adequate approximation to the tail of the data. The pairwise likelihood method was found to be more robust to model misspecification.

We modelled rainfall extremes using a two-step approach that combines a latent variable model for the margins with the fit of an elliptical Pareto process. The Bayesian approach enables the modelling of complex spatial trends in univariate marginal distributions and provides a more flexible alternative to regression models. Ideally we would like to use a full Bayesian model based on generalized elliptical Pareto processes, but it is prone to computational difficulties. In our application, one evaluation of the censored likelihood function takes several minutes, preventing the use of this likelihood in Markov chain Monte Carlo algorithms.

The distribution of an elliptical Pareto process is fully specified by its exponent measure $\Lambda_+$ which depends on a correlation function and a shape parameter. In practice, data are observed on a finite set of sites $s$, so inference focuses on the estimation of the finite-dimensional exponent measure $\Lambda_{+,s}$, based on the choice of a risk functional that can be evaluated at $s$. Estimation is based on finite-dimensional distributions from which parameters can be identified, which induces the unique dependence structure on $C_+(K)\setminus\{0\}$. It thus enables projection of extremes on $C_+(K)\setminus\{0\}$, eventually using different risk functionals than those used for the estimation.

The properties of the maximum likelihood estimators for elliptical $\ell$-Pareto processes under the limiting model are well-known but further work to investigate the theoretical properties of such estimators under domain of attraction assumptions would be valuable.

\section*{Acknowledgments}
We thank A.~C.~Davison, R.~Huser and two anonymous referees for their comments on the manuscript. We thank the Swiss National Science Foundation for financial support. T. Opitz is grateful for financial support from two projects:  McSim, funded by the French National Research Agency,  and MIRACCLE, funded by the French Ministry for ecology.

\section*{Supplementary material}
Supplementary material available at Biometrika online includes details on the latent variable model used in Section~\ref{sec:appli} and on a new estimator for extremal coefficients, and discussion on asymptotic independence for the rainfall data.

%\bibliographystyle{customE}
%\bibliography{Extremal-t}

\end{document}